\documentclass[a4paper,twoside]{article}

\usepackage{placeins}
\usepackage{epsfig}
\usepackage{subfigure}
\usepackage{calc}
\usepackage{amssymb}
\usepackage{amstext}
\usepackage{amsmath}
\usepackage{amsthm}
\usepackage{multicol}
\usepackage{pslatex}
\usepackage{apalike}
\usepackage{color}
\usepackage{listings}
\usepackage{tikz,ifthen}
\usepackage{hyperref}
\usepackage{paralist}
\usepackage{xspace}
\usepackage{SCITEPRESS}     

\usepackage{tikz,ifthen,pgfplots}
\usetikzlibrary{arrows,trees,backgrounds,automata,shapes,decorations,plotmarks,fit,calc,positioning,shadows,chains}
\tikzstyle{every pin edge}=[<-,shorten <=1pt]
\tikzstyle{neuron}=[circle,fill=black!25,minimum size=17pt,inner sep=0pt]
\tikzstyle{input neuron}=[neuron, fill=green!50]
\tikzstyle{output neuron}=[neuron, fill=red!50]
\tikzstyle{hidden neuron}=[neuron, fill=blue!50]
\tikzstyle{annot} = [text width=6em, text centered]
\usetikzlibrary{calc}

\DeclareFixedFont{\ttb}{T1}{txtt}{bx}{n}{7.5} 
\DeclareFixedFont{\ttm}{T1}{txtt}{m}{n}{7.5}  

\newcommand{\relu}{\text{ReLU}\xspace{}}

\definecolor{deepblue}{rgb}{0,0,0.5}
\definecolor{deepred}{rgb}{0.6,0,0}
\definecolor{deepgreen}{rgb}{0,0.5,0}

\newcommand{\ra}{\rangle}

\newcommand{\compose}{\parallel}

\newcommand{\request}{{\color{blue}request}}
\newcommand{\waitfor}{{\color{green!50!black}wait for}}
\newcommand{\blocking}{{\color{red}blocking}}

\renewcommand\j[1]{\textsc{#1}}

\begin{document}

\title{Guarded Deep Learning using Scenario-Based Modeling}

\author{\authorname{Guy Katz}
 \affiliation{The Hebrew University of Jerusalem, Jerusalem, Israel}
 \email{guykatz@cs.huji.ac.il}}


\abstract{ Deep neural networks (DNNs) are becoming prevalent,
  often outperforming manually-created systems. Unfortunately, DNN
  models are opaque to humans, and may behave in unexpected ways when
  deployed. One approach for allowing safer deployment of DNN models
  calls for augmenting them with hand-crafted \emph{override rules},
  which serve to override decisions made by the DNN model when certain
  criteria are met.  Here, we propose to bring together DNNs and the
  well-studied \emph{scenario-based modeling} paradigm, by expressing
  these override rules as simple and intuitive scenarios. This
  approach can lead to override rules that are comprehensible to
  humans, but are also sufficiently expressive and powerful to
  increase the overall safety of the model. We describe how to extend
  and apply  scenario-based modeling to this new setting, 
  and demonstrate our proposed technique on multiple
  DNN models.}

\onecolumn \maketitle \normalsize \vfill

\section{\uppercase{Introduction}}
\label{sec:introduction}
\noindent

Deep machine learning~\cite{FoBeCu16} is dramatically changing our
world, by allowing engineers to create complex models using
automated learning algorithms~\cite{GoSoTaCaRiBaAmTeMa18}.
These learning algorithms generalize examples of how the desired system should behave into an artifact called a \emph{deep neural
  network} (\emph{DNN}), capable of correctly handling new inputs --- even if it had not
encountered them previously.  In many instances, DNNs have been shown to \emph{greatly
  outperform} manually-crafted software. Examples of note include
AlphaGO~\cite{SiHuMaGuSiVaScAnPaLaDi16}, which defeated some of the
world's strongest human Go players; DNN-based systems for image
recognition with super-human precision~\cite{SiZi14}; and 
systems in many other domains such as natural language
processing~\cite{CoWeBoKaKaKu11}, recommender systems~\cite{ElSoHe15}
and bioinformatics~\cite{ChSaBa14}.  
As DNNs
are proving more accurate and easier to create than
manually-crafted systems, their use is expected to continue and
intensify in the coming decades. Indeed,
there is now even a trend of using DNNs
in \emph{highly critical systems}, such as autonomous cars and unmanned
aircraft~\cite{BoDeDwFiFlGoJaMoMuZhZhZhZi16,JuLoBrOwKo16}.

DNNs have been demonstrating extraordinary performance, but they also pose
new challenges~\cite{AmOlStChScMa16}. A key difficulty is that DNNs
are extremely \emph{opaque}: because they are generated by computers
and not by humans, we can empirically see that they perform well, but
we do not fully understand their internal decision
making. Consequently, it is nearly impossible for humans to reason
about the correctness of DNNs. For instance, it has been observed that
many state-of-the-art DNNs for image recognition, which at first
glance seemed to perform spectacularly, could be fooled by slight
perturbations to their inputs~\cite{SzZaSuBrErGoFe13}. This raises
serious concerns about these networks' safety and reliability.
Initial attempts are being made to automatically reason about DNNs 
using formal methods~\cite{KaBaDiJuKo17,HuKwWaWu16,GeMiDrTsChVe18,WaPeWhYaJa18,Marabou2019}, but
these approaches are currently of limited scalability. Further, these
approaches do not specify how to correct an undesirable behavior in a DNN after
it has been discovered, which is also a difficult task.

Consider, for example, the case of the DeepRM
system~\cite{MaAlMeKa16DeepRM}. The goal of this system is to perform
resource allocation: the system has available resources (e.g., CPUs and
memory), and a queue of pending jobs; and it needs to either schedule
a pending job and assign some of the resources to it, or perform a
``pass'' action, in which no new jobs are assigned resources and the system
waits for executing jobs to terminate and free up their assigned resources. The
goal is to schedule jobs in a way that maximizes throughput. 
DeepRM achieves this by maintaining a model of the system (resources,
incoming jobs), and using a pre-trained DNN to choose which action to
perform. DeepRM performs very well when compared to
state-of-the-art, manually created software that tackles the same
problem~\cite{MaAlMeKa16DeepRM}.

Despite its overall satisfactory performance, the authors of DeepRM
report that the system may sometimes behave in undesirable ways. For
example, the controller might request that job $x$ be allocated
resources, although no job $x$ exists in the job queue. In their
implementation~\cite{MaAlMeKa16DeepRMCode}, the authors address this
situation by introducing an \emph{override rule}: a piece of code that
examines the current state of the system, and overrides the DNN's
decision when this particular case is detected. Here, the override
rule changes the controller's selection to ``pass'' whenever the
controller requests to allocate resources to a non-existent job.
There are additional override rules included in
DeepRM~\cite{MaAlMeKa16DeepRMCode}, and also in other systems
(e.g., the Pensieve system~\cite{MaNeAl17Pensieve}). Moreover, since DeepRM's release, additional
undesirable behaviors have been discovered~\cite{KaBaKaSc19}, and
addressing these might require augmenting the system with yet additional
override rules in the future.

These cases, and others, indicate that override rules are becoming an
integral component of DNN-based models. As erroneous
behaviors may be discovered after the initial deployment phase,
override rules may need to be added, extended, enhanced and refactored
throughout the system's lifetime. We argue that this situation calls for
leveraging suitable modeling techniques, in a way that will facilitate
creating and maintaining override rules --- leading to overall
increased system reliability.

In this paper we advocate the use of the \emph{scenario-based modeling}
(\emph{SBM}) framework~\cite{HaMaWe12ACM,DaHa01} for creating override
rules. In SBM, individual system behaviors are modeled as independent
scenarios, and are then automatically interwoven when the model is
executed in order to produce cohesive system behavior. SBM has been
shown to afford several benefits in system design and automated
maintenance, and is particularly suitable for \emph{incremental
  development} --- which is a highly desirable trait when dealing with
override rules. We propose here a method for applying SBM to systems
with DNN components, in a way that allows to specify override rules as
SBM scenarios. We discuss the benefits of the approach (in particular,
those afforded by the amenability of SBM to automated
analysis~\cite{HaKaMaWe15}), and demonstrate its applicability to a
few recently proposed systems. Although our focus here is on systems with DNN
components, our approach could be extended to systems with different kinds
of opaque components.

The rest of this paper is organized as follows. In
Section~\ref{sec:background} we provide the necessary background on
SBM, DNNs and override rules. Next, in Section~\ref{sec:scenariosAndDnns}
we present our method for applying SBM to systems with DNN
components. In
Section~\ref{sec:evaluation} we describe an evaluation of our
approach, followed by a discussion of related work in
Section~\ref{sec:relatedWork}.  We conclude in
Section~\ref{sec:conclusion}.

\section{\uppercase{Background}}
\label{sec:background}
\noindent

\subsection{Deep Neural Networks and Override Rules}
\label{sec:background:dnns}
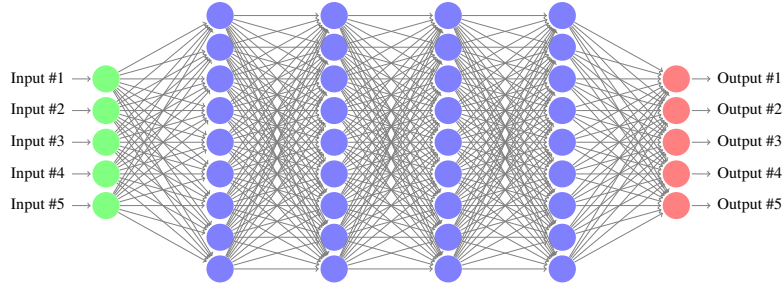
\begin{figure*}
\begin{center}
\scalebox{0.6}{
\def\layersep{2.5cm}
\def\vertSepFactory{0.7}
\begin{tikzpicture}[shorten >=1pt,->,draw=black!50, node distance=\layersep]
    \foreach \name / \y in {1,...,5}
        \node[input neuron, pin=left:Input \#\y] (I-\name) at (0,-\vertSepFactory * \y) {};

    \foreach \name / \y in {1,...,9}
        \path[yshift=1.4cm]
            node[hidden neuron] (H1-\name) at (1*\layersep,-\vertSepFactory * \y cm) {};

    \foreach \name / \y in {1,...,9}
        \path[yshift=1.4cm]
            node[hidden neuron] (H2-\name) at (2*\layersep,-\vertSepFactory * \y cm) {};

    \foreach \name / \y in {1,...,9}
        \path[yshift=1.4cm]
            node[hidden neuron] (H3-\name) at (3*\layersep,-\vertSepFactory * \y cm) {};

    \foreach \name / \y in {1,...,9}
        \path[yshift=1.4cm]
            node[hidden neuron] (H4-\name) at (4*\layersep,-\vertSepFactory * \y cm) {};

    \foreach \name / \y in {1,...,5}
        \node[output neuron,pin={[pin edge={->}]right:Output \#\y}]
        (O-\name) at (5*\layersep, -\vertSepFactory * \y cm) {};

    \foreach \source in {1,...,5}
        \foreach \dest in {1,...,9}
            \path (I-\source) edge (H1-\dest);

    \foreach \source in {1,...,9}
        \foreach \dest in {1,...,9}
            \path (H1-\source) edge (H2-\dest);

    \foreach \source in {1,...,9}
        \foreach \dest in {1,...,9}
            \path (H2-\source) edge (H3-\dest);

    \foreach \source in {1,...,9}
        \foreach \dest in {1,...,9}
            \path (H3-\source) edge (H4-\dest);

    \foreach \source in {1,...,9}
        \foreach \dest in {1,...,5}
            \path (H4-\source) edge (O-\dest);

\end{tikzpicture}
}
\caption{A fully connected DNN with 5 input nodes (in green), 5 output
  nodes (in red), and
  4 hidden layers containing a total of 36 hidden nodes (in blue).
}
\label{fig:fullyConnectedNetwork}
\end{center}
\end{figure*}

Deep neural networks (DNNs) are directed graphs, in which the nodes
(neurons) are organized into layers. The first layer is the input
layer, the last layer is the output layer, and the multiple remaining
layers are the hidden layers. Each node in the network (except for
input nodes) is connected to nodes from the preceding layer, using
predetermined weight values (an illustration appears in
Fig.~\ref{fig:fullyConnectedNetwork}). Selecting appropriate weight
values is key, and is performed during a \emph{training} phase, which
is beyond the scope of this paper (for a survey, see,
e.g.,~\cite{FoBeCu16}). A DNN is evaluated by assigning values to its
input neurons, and then propagating these values forward through the
network, each time computing values for a given layer from the values
of its predecessor. Eventually, the output values (i.e., the values of
neurons in the output layer) are computed, and
are returned to the user. Often, DNNs are used as controllers or
classifiers, in which case it is typical to return to the user the
index of the output neuron that received the highest value. This neuron indicates the
action, or classification, determined by the DNN.

For our purpose here, it is enough to regard a DNN as a black box,
that transforms an input into an output. However, for completeness, we
briefly describe how a DNN is evaluated. The value of each hidden node
in the network is computed by calculating a weighted sum of the node
values from the previous layer, according to the edge weights. Then, a
non-linear \emph{activation function} is applied to this weighted
sum~\cite{FoBeCu16}, and its result becomes the value of the node
being computed.  For simplicity we focus here on the Rectified Linear
Unit (ReLU) activation function~\cite{NaHi10}, given by
$\relu{}(x) = \max{}(0, x)$.  Thus, when a node uses the ReLU
activation function, its value is calculated as the maximum of the
linear combination of nodes from the previous layer and $0$.

\begin{figure}[htp]
  \begin{center}
    \scalebox{1} {
      \def\layersep{2.5cm}
    \begin{tikzpicture}[shorten >=1pt,->,draw=black!50, node distance=\layersep,font=\footnotesize]

      \node[input neuron] (I-1) at (0,-1cm) {$x_1$};
      \node[input neuron] (I-2) at (0,-2cm) {$x_2$};

      \path[yshift=0.5cm] node[hidden neuron] (H-1)
      at (\layersep,-1cm) {$v_1$};
      \path[yshift=0.5cm] node[hidden neuron] (H-2)
      at (\layersep,-2cm) {$v_2$};
      \path[yshift=0.5cm] node[hidden neuron] (H-3)
      at (\layersep,-3cm) {$v_3$};

      \node[output neuron] at (2*\layersep, -1) (O-1) {$y_1$};
      \node[output neuron] at (2*\layersep, -2) (O-2) {$y_2$};

      \path (I-1) edge[] node[above,pos=0.4] {$1$} (H-1);
      \path (I-1) edge[] node[above,pos=0.4] {$-2$} (H-2);
      \path (I-2) edge[] node[below,pos=0.3] {$-1$} (H-1);
      \path (I-2) edge[] node[below,pos=0.3] {$3$} (H-3);

      \path (H-1) edge[] node[above] {$1$} (O-1);
      \path (H-2) edge[] node[above] {$-2$} (O-1);
      \path (H-2) edge[] node[above] {$1$} (O-2);
      \path (H-3) edge[] node[above] {$1$} (O-2);

      \node[annot,above of=H-1, node distance=1cm] (hl) {Hidden layer};
      \node[annot,left of=hl] {Input layer};
      \node[annot,right of=hl] {Output layer};
    \end{tikzpicture}
    }
    \captionsetup{size=small}
    \captionof{figure}{A small neural network with a single hidden layer.}
    \label{fig:runningExample}
  \end{center}
\end{figure}
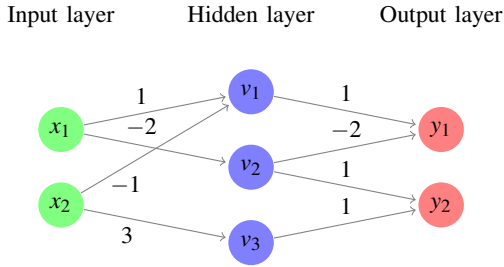

Fig.~\ref{fig:runningExample} depicts a small DNN that will serve as a
running example. The network acts as a controller: it has two
inputs, $x_1$ and $x_2$; three hidden neurons, $v_1,v_2$ and $v_3$,
each with the ReLU activation functions; and it selects one of two
possible actions through its output neurons, $y_1$ and $y_2$. We
slightly abuse notation, and use $y_1$ and $y_2$ to denote both the
neurons and the actions/classes those neurons represent. The selected
action is the one assigned the highest score. We see, for example,
that assigning $x_1=1, x_2=0$ results in output values $y_1=1, y_2=0$,
i.e., action $y_1$ is selected; whereas $x_1=0, x_2=1$ leads to
$y_1=0, y_2=3$, i.e. action $y_2$ is selected.

An \emph{override rule} is a triple $\langle P, Q, \alpha\rangle$, where $P$
is a predicate over the network's
inputs, $Q$ is a predicate over the network's outputs, and $\alpha$ is an override
action. The semantics of these rules is that if $P$ and $Q$ hold for a
network's evaluation, then output action $\alpha$ should be selected ---
regardless of the network's output. For example, we might specify the rule
\[ 
  \langle x_1 > 0 \wedge x_2 < x_1,  true,  y_2 \rangle
\]
which would be triggered for inputs $x_1=1, x_2=0$. As we saw
previously, in this case the network outputs $y_1$; but with the
override rule, this selection would be changed to $y_2$.
By setting $Q$ to true, we created an override rule the only examines
the DNN's inputs. We could, for example, set $Q$ to $y_2>10$, in which
case the rule would not be triggered for $x_1=1, x_2=0$. 
By adjusting $P$ and $Q$, the aforementioned formulation can express
many common override rules, such as the
rule in the DeepRM example described in Section~\ref{sec:introduction}.

\subsection{Scenario-Based Modeling}
Scenario-based modeling~\cite{HaMaWe12ACM} is an approach for modeling
complex reactive systems. At the core of the approach lies the notion
of a \emph{scenario object}: a description of a single behavior, whether desirable
or undesirable, of the system being modeled. Each scenario object is
created separately, and does not directly interact with the other
scenarios; instead, it only interacts with a global execution
mechanism. This execution mechanism can execute a set of scenarios in
a way that produces cohesive, global behavior.

There are several flavors of scenario-based modeling, which may differ
in the various idioms that a scenario object uses to interact with
the execution mechanism and affect the overall execution of the
system. Here, we focus on the commonly used idioms of
\emph{requesting}, \emph{waiting-for} and \emph{blocking} events~\cite{HaMaWe12ACM}. When
executed, each scenario object may declare it has reached a
\emph{synchronization point}, in which the execution infrastructure
must trigger an event. The object then specifies which events it would
like to have triggered (\emph{requested} events); which events it
forbids from being triggered (\emph{blocked} events); and which events
it does not actively request, but should be notified in case
they are triggered by the execution mechanism (\emph{waited-for events}). The execution
infrastructure waits for all the scenario objects to synchronize (or
just for a subset thereof, depending on the semantics used~\cite{HaKaKa13}); selects
an event that is requested and not blocked for triggering; and informs
any relevant scenario object that this event has been triggered.

\begin{figure}[htp]
  \centering
  \scalebox{0.65} {
    
    \tikzstyle{box}=[draw,  text width=2cm,text centered,inner sep=3]
    \tikzstyle{set}=[text centered, text width = 10em]

    \begin{tikzpicture}[thick,auto,>=latex',line/.style ={draw, thick, -latex', shorten >=0pt}]
      
      \matrix(bts) [row sep=0.3cm,column sep=2cm]  {

        \node (box1)  [box] {\waitfor{} \j{WaterLow}}; \\
        \node (box2)  [box] {\request\ \j{AddHot}}; \\
        \node (box3)  [box] {\request\ \j{AddHot}}; \\ 
        \node (box4)  [box] {\request\ \j{AddHot}}; \\ 
      };

      \draw [->] ($(box1.north) + (0,0.3cm)$) -- (box1.north);
      \node (title) [above=0.1cm of bts,box,draw=none] at ($(bts) + (-0.25cm,2.51cm)$) 
      {\j{AddHotWater}};  
      
      \begin{scope}[every path/.style=line]
        \path (box1)   -- (box2);
        \path (box2)   -- (box3);
        \path (box3)   -- (box4);
        \path (box4.east)   -- +(.25,0) |- (box1);
      \end{scope}

      \matrix(bts2) [right=.25cm of bts, row sep=0.3cm,column sep=2cm] {
        \node (box1)  [box] {\waitfor{} \j{WaterLow}}; \\
        \node (box2)  [box] {\request\ \j{AddCold}}; \\
        \node (box3)  [box] {\request\ \j{AddCold}}; \\ 
        \node (box4)  [box] {\request\ \j{AddCold}}; \\ 
      };
      
      \draw [->] ($(box1.north) + (0,0.3cm)$) -- (box1.north);
      \node (title) [above=0.1cm of bts2,box,draw=none] at ($(bts2) + (-0.25cm,2.51cm)$) 
      {\j{AddColdWater}};

      \begin{scope}[every path/.style=line]
        \path (box1)   -- (box2);
        \path (box2)   -- (box3);
        \path (box3)   -- (box4);
        \path (box4.east)   -- +(.25,0) |- (box1);
      \end{scope}

      \matrix(bts3) [right=.25cm of bts2, row sep=0.3cm,column sep=2cm] {
        \node (box1)  [box] {\waitfor{}  \j{AddHot} while  \blocking\ \j{AddCold}}; \\
        \node (box2)  [box] {\waitfor{}  \j{AddCold} while \blocking\ \j{AddHot}}; \\
      };

      \draw [->] ($(box1.north) + (0,0.3cm)$) -- (box1.north);
      \node (title) at (title-|bts3) [box,draw=none] {\j{Stability}};  

      \begin{scope}[every path/.style=line]
        \path (box1)   -- (box2);
        \path (box2.east)   -- +(.25,0) |- (box1);
      \end{scope}
      
      \node (log)  [right=.3cm of bts3,box,text width=2cm,fill=yellow!20] {
        $\cdots$ \\ 
        \j{WaterLow} \\
        \j{AddHot}  \\ 
        \j{AddCold} \\ 
        \j{AddHot}  \\ 
        \j{AddCold} \\ 
        \j{AddHot}  \\ 
        \j{AddCold} \\ 
        $\cdots$ \\
      }; 

      \node (title2) at (title-|log)            
      [box,draw=none] {\j{Event Log}};  
    \end{tikzpicture}
  }  
  \caption{
    (From~\cite{HaKaMaWe14})
    A scenario-based model of a
    system that controls the water level in a tank with hot and
    cold water taps. 
  }  
  \label{fig:watertap}
\end{figure}
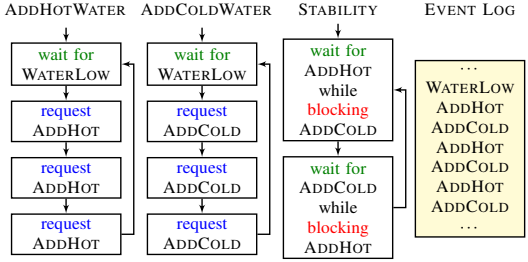

A toy example of a scenario-based model appears in
Fig.~\ref{fig:watertap}. The model depicted therein belongs to a system
that controls the water level in a tank with hot and
cold water taps. Each scenario object is depicted as a transition
system, where the nodes represent the predetermined synchronization points.
The scenario object
\j{AddHotWater} repeatedly waits for \j{WaterLow} events and requests three
times the event \j{AddHot}; and the 
scenario object \j{AddColdWater} performs a symmetrical operation
with cold water. In a model that includes only the objects \j{AddHotWater} and
\j{AddColdWater}, the three \j{AddHot} events and three
\j{AddCold} events may be triggered in any order during execution.
In order to maintain the stability of the water temperature in the tank,
the scenario object
\j{Stability} enforces the interleaving of \j{AddHot} and \j{AddCold} events 
by using event blocking. The execution trace of the resulting model is
depicted in the event log.

SBM has been implemented in a variety of high-level languages, such as
Java~\cite{HaMaWe10BPJ}, C++~\cite{HaKa14}, JavsScript~\cite{BaWeRe18}
and ScenarioTools~\cite{GeGrGuKoGlMaKa17}. The methodology has been
successfully used in modeling complex systems, such as
web-servers~\cite{HaKa14}, cache coherence protocols~\cite{HaKaMaMa16}
and robotic controllers~\cite{GrGr18b}. For simplicity and generality,
in the remainder of this paper we mostly describe scenario-based models in
terms of transitions systems.

Following the definitions in~\cite{Ka13}, we formalize the
SBM framework as follows. A scenario object
$O$ over event set $E$ is a tuple
$O = \langle Q, \delta, q_0, R, B \ra$, where the components are
interpreted as follows:
\begin{itemize}
\item $Q$ is a set of states, each representing one of the
  predetermined synchronization points;
\item $q_0$ is the initial state;
\item $R:Q\to 2^E$ and $B:Q\to 2^E$
map states to the sets of events requested and blocked at these
states (respectively); and
\item $\delta: Q \times E \to 2^Q$ is a transition function,
  indicating how the object reacts when an event is triggered.
  \end{itemize}

Scenario objects can be composed into a single, larger scenario
object, as follows.
For objects
 $O^1 = \langle Q^1, \delta^1, q_0^1, R^1, B^1 \ra$ and
 $O^2 = \langle Q^2, \delta^2, q_0^2, R^2, B^2 \ra$ over a
 common event set $E$, we define the composite scenario object
 $O^1\compose O^2$ as
$O^1 \compose O^2 = \langle Q^1 \times Q^2, \delta,
\langle q_0^1,q_0^2\rangle, R^1\cup R^2, B^1\cup B^2\rangle
$,
where:
\begin{itemize}
\item $\langle \tilde{q}^1,\tilde{q}^2\rangle \in  \delta(\langle q^1,q^2\rangle, e)$
if and only if $\tilde{q}^1 \in \delta^1(q^1,e)$ and $\tilde{q}^2\in
\delta^2(q^2,e)$; and
\item The union of the labeling functions is defined in the natural way; e.g. $e\in (R^1\cup
R^2)(\langle q^1,q^2 \rangle)$ if and only if $e \in R^1(q^1) \cup
R^2(q^2)$, and
$e\in (B^1\cup
B^2)(\langle q^1,q^2 \rangle)$ if and only if $e \in B^1(q^1) \cup
B^2(q^2)$.
\end{itemize}

A \emph{behavioral model} $M$ is defined as a collection of scenario
objects  $O^1, O^2,\ldots, O^n$,
and the executions of $M$ are the executions of the composite object
$O = O^1\compose O^2\compose\ldots\compose O^n$.
Each such execution starts from the initial state of $O$,
and in each state $q$ along the run an enabled event is chosen for triggering, if
one exists (i.e., an event $e\in R(q) - B(q)$).
 Then, the execution moves to a state $\tilde{q}\in \delta(q,e)$, and
so on. 

One extension of SBM, which will be useful in our context, is to treat
events as \emph{variables}~\cite{KaMaSaWe19}. For example, an event
$e$ can be declared to be of type integer. Then, one scenario object
might \emph{request} $e\geq 5$, while another object might block
$e\geq 7$. The execution framework would then employ a
\emph{constraint solver}, such as an SMT solver~\cite{BaTi18}, to
resolve the constraints and trigger, e.g., the event $e=6$. We omit
here the formal definition of this extension, which is
straightforward, and refer the interested reader to~\cite{KaMaSaWe19}.

\section{\uppercase{Modeling Override Scenarios}}
\label{sec:scenariosAndDnns}
\noindent

In the DeepRM case, override rules have been added as unrestricted Python code
within the module that invokes the DNN and processes its
result~\cite{MaAlMeKa16DeepRMCode}. Thus, while the DNN controller
itself is clearly structured and well defined,
override rules are phrased as arbitrary pieces of code. This
could lead to several complications:
\begin{inparaenum}[(i)]
\item as the number of override rules increases, they might become
  convoluted and difficult to comprehend, extend and maintain;
\item the semantics of override rules might be unclear. For example,
  in the case of multiple rules that can all be applied, which one
  prevails?  Is there a particular order in which they should be
  checked? Can rules interact? etc; and
\item the conditions employed within these override rules might become more
  complex, hiding away some of the model's logic where other developers
  might not expect to find it.
\end{inparaenum}

Here, we propose to model override rules using SBM, as a means for
mitigating these difficulties. SBM is geared towards incremental
modeling, which seems a particularly likely scenario when DNNs are
involved: due to the opacity of DNNs, some undesirable behaviors are
likely to be detected only after deployment, requiring the addition of new override
rules. Further, SBM's simple semantics would guarantee that interactions
between the override rules are well defined. Finally, there is a
substantial body of work on automatically verifying, analyzing and
optimizing SBM models, which could prove useful in detecting conflicts
between override rules or simplifying them when their number increases.

\subsection{Modeling DNNs and Override Rules in SBM}

\begin{figure}[ht]
\centering
  \scalebox{0.65} {
    
    \tikzstyle{box}=[draw,  text width=3cm,text centered,inner sep=3]
    \tikzstyle{set}=[text centered, text width = 10em]

    \begin{tikzpicture}[thick,auto,>=latex',line/.style ={draw, thick, -latex', shorten >=0pt}]

      \node (box1)  [initial,box] {{\color{white} hidden text hidden}
        wait for $e_{i_1}, e_{i_2}$ {\color{white} hidden text hidden}};

      \node (box2)  [box, right = 3cm of box1] {request
        $e_{y_1},e_{y_2}$ and block all other events};

      \path[->] (box1) edge [bend right,thick] node {$e_{i_1}, e_{i_2}$} (box2);
      \path[->] (box2) edge [bend right,thick] node[swap] {$e_{y_1},e_{y_2}$} (box1);
      
    \end{tikzpicture}
  }  
\caption{A scenario $O_{DNN}$ for the neural network in
  Fig.~\ref{fig:runningExample}. Events $e_{i_1}$ and $e_{i_2}$
  represent the inputs to the neural network, and events $e_{y_1}$ and
$e_{y_2}$ represent its outputs.}
  \label{fig:scenarioObject}
\end{figure}
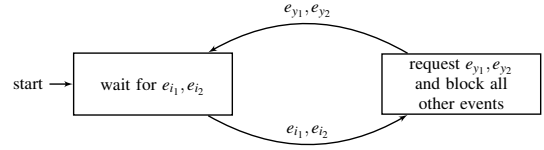

We propose the following method for creating SBM models that combine
 scenario objects and a DNN controller. The core idea is to
represent the DNN as a dedicated scenario object, $O_{DNN}$, to be
included in the scenario-based model. This $O_{DNN}$ is a
non-deterministic scenario that models the DNN controller, thus
allowing it to interact with the other scenario objects. Let us
assume, for the sake of simplicity (we relax this limitation later),
that there is a finite set of possible inputs to the DNN, denoted
$\mathbb{I}$; and let $\mathbb{O}$ denote the set of possible
actions among which the DNN chooses. We introduce new events to
our event set $E$: an event $e_i$ for every $i\in \mathbb{I}$, and an
event $e_o$ for every $o\in \mathbb{O}$. We have our new scenario
object $O_{DNN}$ repeatedly wait for all events $e_i$, and then
request all events $e_o$. This behavior represents the black-box
nature of the DNN, as far as the rest of the model is concerned: we
only know that after an input arrives, one of the outputs will be
selected, without knowing which.  However, when the model is executed,
the execution infrastructure resolves this non-determinism by running the actual DNN and
triggering the output event that corresponds to its selection. For
example, assuming just two possible inputs, e.g. $i_1 = \langle
1,0\rangle$ and $i_2 = \langle 0,1\rangle$, the network depicted in Fig.~\ref{fig:runningExample} would
be represented by the scenario object described in
Fig.~\ref{fig:scenarioObject}.

We introduce the convention that other scenario objects in the systems
may wait-for, but may not block, the input events $e_i$. A single
dedicated scenario, called a \emph{sensor}, is responsible for
requesting an input event when the DNN needs to be evaluated (e.g.,
following a user action). By another convention, no scenario object
except $O_{DNN}$ may request any of the output events $e_o$; however,
other scenario objects may wait-for or block these events. During
execution, if the DNN assigns the highest score to an event that is
currently blocked, we resolve the non-determinism of $O_{DNN}$ by
selecting the event representing the output with the next-to-highest
score, and so on. If no events are left unblocked, then the system is
deadlocked and the execution terminates.

The motivation underlying our definitions is to allow scenario objects
to monitor the inputs and outputs of the DNN controller, by waiting
for their respective events; and then to interfere with the
recommendation of the DNN, by blocking certain output events from
being triggered, which is the main use-case of override rules. Note that a
scenario object may force the DNN to produce a specific output, by blocking all
other possibilities; or it may interfere more subtly, by blocking some
events and allowing the DNN to choose among the remaining events.

In practice, our assumption that the sets $\mathbb{I}$ of possible DNN
inputs and the set $\mathbb{O}$ of possible DNN outputs are finite might be a
limiting factor: for example, in the override rule described in
Section~\ref{sec:background:dnns}, the relative assignments to $x_1$
and $x_2$ affected whether the rule could be triggered or not, and so
it is important to express in our model the exact assignments for
$x_1$ and $x_2$. Of course, there are infinitely many possible assignments.
To waive this limitation we again turn to the extension to
SBM~\cite{KaMaSaWe19} that allows us to treat events as variables of
certain types. We change our formulation slightly: scenario objects in the
system may wait-for a single, composite event that indicates that values have been
assigned to (all of) the DNN's inputs or outputs, and may then act according to
those values.

Using this extension, the override rule from
Section~\ref{sec:background:dnns} is expressed as a scenario object in
Fig.~\ref{fig:overrideRuleAsThread}. The scenario 
enforces the override rule that whenever 
 $x_1 > 0$ and $x_2 < x_1$, output event $y_2$ (and not $y_1$) should
 be triggered. Here, $\langle e_{x_1}, e_{x_2} \rangle$ represents a
 single event that indicates that values have been assigned to the
 DNN's inputs. This event contains two real values, $x_1$ and $x_2$,
 that the scenario can then access and use to determine its
 transition. Event $e_{y_1}$ indicates, as before, that the scenario
 forbids the DNN from selecting $y_1$ as its output action.

\begin{figure}[ht]
\centering
  \scalebox{0.65} {
    
    \tikzstyle{box}=[draw,  text width=3cm,text centered,inner sep=3]
    \tikzstyle{set}=[text centered, text width = 10em]

    \begin{tikzpicture}[thick,auto,>=latex',line/.style ={draw, thick, -latex', shorten >=0pt}]

      \node (box1)  [initial,box] {wait for $\langle x_1, x_2\rangle$};

      \node (box2)  [box, right = 3cm of box1] {block $e_{y_1}$};

      \path[->] (box1) edge [bend right,thick] node[swap] {$x_1>0 \wedge x_2<x_1$} (box2);
      \path[->] (box2) edge [bend right,thick] node[swap] {$e_{y_2}$} (box1);

      \path[] (box1) edge [loop above,thick] node {$x_1\leq 0 \vee
        x_2\geq x_1$} (box1);
      
    \end{tikzpicture}
  }  
  \caption{A scenario object enforcing the override rule that whenever 
 $x_1 > 0$ and $x_2 < x_1$, output event $y_2$ should
 be triggered.
  }
  \label{fig:overrideRuleAsThread}
\end{figure}

\subsection{Liveness Properties}
\label{sec:livenessProperties}
Override rules are typically used to enforce safety properties (``bad
things never happen''). However, there is sometimes a need to enforce
\emph{liveness} properties (``good things eventually happen''). In
particular, this can happen in the context of online reinforcement
learning~\cite{rl-book} --- where the DNN controller might change over
time, and we wish to ensure that it eventually tries out new actions. If
these actions turn out to be beneficial, the RL mechanism will ensure
that the DNN repeats them in the future. Liveness properties are
also relevant when there are fairness constraints; for example, if we
wish to ensure that in a resource allocation system, every pending job
eventually gets scheduled.

An example appears in~\cite{KaBaKaSc19}, where the authors discuss the
\emph{Custard} system: a congestion control system that uses a DNN to
monitor the conditions of a computer network and select a sending bit
rate, in order to minimize congestion~\cite{JaRoBrScTa18}.
In~\cite{KaBaKaSc19}, Custard is examined to see if there are cases in
which the DNN controller chooses a sub-optimal sending rate that does
not utilize all available bandwidth, and never attempts to increase
this bit rate. This kind of behavior constitutes a liveness violation,
which we would like to prevent using an override rule.

SBM can encode the fact that one or multiple DNN output actions should
eventually be blocked, thus forcing the DNN controller to pick a
different alternative. This can be enforced by having a scenario
object wait for $n$ consecutive rounds where a particular output is
triggered, and then block it; an example for $n=3$ appears in
Fig.~\ref{fig:livenessOverrideRule}, where a scenario looks for 3
consecutive DNN evaluations where $y_2$ is triggered, after which it
blocks $y_2$ once, forcing the DNN to select another action.  An
alternative is to have the override rule block the particular output
event with a very low probability~\cite{HaKaMaWe14}, thus enforcing
the fact that it will \emph{eventually} be blocked with probability
$1$.

\begin{figure}[ht]
\centering
  \scalebox{0.65} {
    
    \tikzstyle{box}=[draw,  text width=1.7cm,text centered,inner sep=3]
    \tikzstyle{set}=[text centered, text width = 10em]

    \begin{tikzpicture}[thick,auto,>=latex',line/.style ={draw, thick, -latex', shorten >=0pt}]

      \node (box1)  [initial,box] {wait for $e_{x_1}, e_{x_2}$};
      \node (box2)  [box, below = 1.5cm of box1] {wait for $e_{y_1,y_2}$};

      \node (box3)  [box, right = 1cm of box1] {wait for $e_{x_1}, e_{x_2}$};
      \node (box4)  [box, below = 1.5cm of box3] {wait for $e_{y_1,y_2}$};

      \node (box5)  [box, right = 1cm of box3] {wait for $e_{x_1}, e_{x_2}$};
      \node (box6)  [box, below = 1.5cm of box5] {wait for $e_{y_1,y_2}$};

      \node (box7)  [box, below = 1.5cm of box4] {wait for
        $e_{x_1,x_2}$};
      \node (box8)  [box, left = 1cm of box7] {block $e_{y_2}$, wait
        for $e_{y_1}$};

      \path[->] (box1) edge [bend left,thick] node[] {$*$} (box2);
      \path[->] (box2) edge [bend left,thick] node[] {$e_{y_1}$}
      (box1);
      \path[->] (box2) edge [thick] node[] {$e_{y_2}$} (box3);
 
      \path[->] (box3) edge [bend left,thick] node[swap] {$*$} (box4);
      \path[->] (box4) edge [bend right,thick] node[pos=0.3] {$e_{y_1}$}
      (box1);
      \path[->] (box4) edge [thick] node[] {$e_{y_2}$} (box5);

      \path[->] (box5) edge [bend left,thick] node[swap] {$*$} (box6);
      \draw[->] (box6.east) -| node[swap] {$e_{y_1}$} 
      ($(box5.north east) + (0.3,0.3)$) -|  
      ($(box1.north) + (0,0.3)$) -| (box1.north);
      \path[->] (box6) edge [thick] node[] {$e_{y_2}$} (box7);

      \path[->] (box7) edge [thick] node[swap] {$*$} (box8);

      \draw[->] (box8.west) -| 
      ($(box8.west) + (-0.5,0)$) -- node[] {$e_{y_1}$}
      ($(box1.west) + (-0.5,-0.8)$) -- (box1.south west);
      
    \end{tikzpicture}
  }  
  \caption{A scenario object that enforces a liveness property for the
    network from Fig.~\ref{fig:scenarioObject}.  }
  \label{fig:livenessOverrideRule}
\end{figure}

\subsection{Automated Analysis}

Scenario-based modeling has been shown to facilitate automated formal
analysis~\cite{HaKaMaWe15}. Specifically, the simple synchronization
constructs that scenario objects in SBM models use render tasks such
as model checking~\cite{KaBaHa15}, compositional verification~\cite{HaKaKaMaMiWe13} and automated repair~\cite{Ka13}
simpler than they would be for less restricted models. We argue that
these properties add to the attractiveness of SBM as a formalism for
expressing override rules.

One particular use case that illustrates the aforementioned claim is
deadlock freedom. As additional override rules are added, perhaps by
different modelers, there is a risk that a certain sequence of inputs
to the DNN might cause a deadlock. As a simple illustrative example,
consider the override rule expressed in
Fig.~\ref{fig:overrideRuleAsThread}: whenever $x_1>0$ and $x_2<x_1$,
output $y_2$ should be selected. Suppose now that another modeler,
concerned about the fact that the DNN might always advise $y_2$, adds
the override rule depicted in Fig.~\ref{fig:livenessOverrideRule}:
after $3$ consecutive $y_2$ events, a different event must be
selected. These two override rules together might result in a
deadlock: for example, if the DNN receives the inputs $x_1=2, x_2=1$
three consecutive times, both override rule would be triggered,
simultaneously blocking both output events $e_{y_1}$ and $e_{y_2}$.

Such situations can be avoided by running a verification query that
ensures that the system is deadlock free. This query can be run, e.g.,
after the addition of each new override rule. Should a deadlock be
detected, the counter-example provided by the verification tool could
guide the modeler in changing the conflicting rules --- after which
verification can be run again, to ensure that the system is now
indeed deadlock free. Of course, additional system-specific
properties, beyond deadlock freedom, could also be verified.

\section{\uppercase{Evaluation}}
\label{sec:evaluation}
\noindent

For evaluation purposes, we implemented our approach on top of the BPC
framework for scenario-based modeling in
C++~\cite{HaKa14} (of course, other SBM frameworks could also be
used).
The BPC package allows modelers to leverage many of the powerful
constructs of C++, while forcing them to adhere to the SBM principles:
each scenario is modeled as a separate object, and inter-scenario
interactions are performed through a global execution mechanism that
BPC provides.  Here, we used BPC to model override rules for the
DeepRM system for resource management, and for the Custard system for
congestion control.

\subsection{Override Rules for DeepRM}
The DeepRM system~\cite{MaAlMeKa16DeepRM} (discussed in Section~\ref{sec:introduction})
performs resource allocation: it assigns available resources to
pending jobs, with the goal of maximizing throughput. As part of our
evaluation we implemented an override rule that prevents the DNN
controller from attempting to assign resources to non-existing jobs,
which is an undesirable behavior that occurs in
practice~\cite{MaAlMeKa16DeepRMCode}.

BPC code for our override rule, implemented as a scenario object,
appears in Fig.~\ref{fig:deeprmScenario}. We assume that the queue of
pending jobs is of size 5, and that the DNN's output actions are
denoted $y_0,y_1, \ldots, y_5$.  Action $y_i$ for $1\leq i \leq
5$ means that the job in slot $i$ of the queue should be allocated
resources, and the special action $y_0$ is the ``pass'' action,
indicating that no job should be allocated resources at this time. We
use $x$ to denote an event
indicating that the DNN needs to be evaluated on certain input
values, available as parameters of $x$. The state of the job queue is
part of the input to the DNN controller. Specifically, we use $x[i]$
for $1\leq i \leq 5$ to
denote a Boolean value that indicates whether or not there is currently
a pending job in slot $i$ of the queue.

The override scenario object is implemented as a class that inherits
from BPC's 
special BThread class. The scenario object can then use the special bSync() method to
initialize a synchronization point with the other scenarios in the
model (including $O_{DNN}$, the scenario object that models the DNN controller). This method takes as input three Event
vectors --- the first containing the set of requested events, the
second containing the set of waited-for events, and the third
containing the set of blocked events. The bSync() call suspends the object's execution until
the BPC mechanism selects and triggers an event; then, if the
triggered event was requested or waited-for by the scenario
object, the scenario resumes execution and can retrieve the triggered event
 using the lastEvent() method.

Our scenario object runs in an infinite loop, each time waiting for
the input event $x$ to be triggered. When that happens, it examines
$x$ to determine which slots of the job queue are occupied; and
then synchronizes again to block event $y_i$ for any unoccupied
slots. Note that this scenario object can never cause a deadlock,
because it never blocks event $y_0$.

\begin{figure}[ht]
\begin{lstlisting}
class EnsureJobExists : public BThread {
  void entryPoint() {
    Vector<Event> emptySet = {};
    Vector<Event> allInputs = { x };
    Vector<Event> allOutputs = { $y_0, \ldots, y_5$ };

    while ( true ) {
      bSync( emptySet, allInputs, emptySet );
      lastInput = lastEvent();
      Vector<Event> blocked = {};

      for ( int i = 1; i <= 5; ++i ) {
        if ( !lastInput[i] )
          blocked.append( $y_i$ )
      }

      bSync( emptySet, allOutputs, blocked )
    }
  }
}
\end{lstlisting}
\caption{A scenario object for preventing the DeepRM DNN controller
  from assigning resources to non-existing jobs.}
  \label{fig:deeprmScenario}
\end{figure}

\subsection{Override Rules for Custard}
As briefly discussed in Section~\ref{sec:livenessProperties}, Custard
is a DNN-based congestion control system. The DNN controller takes as
input various readings about the current and previous state of the
computer network (e.g., throughputs, loss rates, and latency), and
selects the next sending bit rate. Custard is a reactive system,
designed to be run continuously and use the results of its past
decisions (as reflected in past network readings) when making its next
choice of bit rate.

Due to the opacity of the DNN controller, one concern when using
Custard is that it might be too \emph{conservative}. Specifically, we
may wish to avoid a situation where the state of the computer network
is completely steady, and yet the DNN controller never tries to
increase the sending bit rate --- and thus never finds out whether
there is additional, currently unused bandwidth.

A scenario object that prevents this case is depicted in
Fig.~\ref{fig:custardScenario}. The scenario looks for a situation
where the DNN's inputs and outputs have been identical for the last
$n=10$ rounds, and when this is detected it blocks the previous output
action. Event $x$ represents here an input assignment (comprised of
multiple input values) on which the DNN has been evaluated, and event
$y$ represents the DNN's output selection. Here, for
simplicity, we do not examine the actual values of $x$, and only look for repeating
assignments (in practice, we may want to apply this override rule
only if the physical network's conditions are both \emph{steady} and
\emph{good}, indicating that there may be unused bandwidth).

\begin{figure}[ht]
\begin{lstlisting}
const int n = 10;

class PreventSteadyState : public BThread {
  void entryPoint() {
    Vector<Event> empty;
    Vector<Event> allInputs = { x };
    Vector<Event> allOutputs = { y };

    Event lastInput;
    Event lastOutput;

    while ( true ) {
      bSync( empty, allInputs, empty );
      lastInput = lastEvent();

      bSync( empty, allOutputs, empty );
      lastOutput = lastOutput();

      bool steadyState = true;
      int i = 1;
      while ( i < n && steadyState ) {
        bSync( empty, allInputs, empty );
        if ( lastInput != lastEvent() )
          steadyState = false;

        bSync( empty, allOutputs, empty );
        if ( lastOutput != lastEvent() )
          steadyState = false;

        ++i;
      }        
 
      if ( steadyState ) {
        bSync( empty, allInputs, empty );
        bSync( empty, allOutputs, lastOuptut );
      }
    }
  }
}
\end{lstlisting}
\caption{A scenario object for enforcing the Custard DNN to choose a
  different action if the state has been steady for $n=10$ iterations.}
  \label{fig:custardScenario}
\end{figure}

\section{\uppercase{Related Work}}
\label{sec:relatedWork}
\noindent

Override rules, sometimes also referred to as \emph{shields}, have
been applied ad-hoc in multiple DNN-enabled systems, such as
DeepRM~\cite{MaAlMeKa16DeepRM} and
Pensieve~\cite{MaNeAl17Pensieve}. Such rules, and related forms of
runtime monitors, are also found in control
systems for robots~\cite{PhYaGrSmSt17}, drones~\cite{DeGhSeShTi18},
and in various other formalisms which are not directed particularly at
deep learning~\cite{HaMoSc06,FaMoFeRi11,ScDeRiGaCoHoStSm15,JiLa17,WuRaRaLaSe18}.  The
formal methods community has recently taken an interest in override
rules for systems with DNNs: for example, by proposing techniques to
synthesize rules that affect the controller as little as
possible~\cite{AvBlChHeKoPr19,WuWaDeWa19}.

Various aspects of the SBM formalism, especially those pertaining to
the formal analysis of scenario-based models, have been studied over
the years. These aspects include the automatic
repair~\cite{HaKaMaWe12}, verification~\cite{HaKaLaMaWe15},
synthesis~\cite{GrGrKaMa16} and
optimization~\cite{HaKaKaMaWeWi15,GrGrKaMaGlGuKo16,StGrGrHaKaMa17,StGrGrHaKaMa18,HaKaMaSaWe20}
of models. SBM is also a key component of the Wise Computing
initiative~\cite{MaArElGoKaLaMaShSzWeHa16,HaKaMaMa16b,HaKaMaMa18},
which seeks to transform the computer into a proactive team member,
capable of developing complex models alongside human engineers.

In this paper we focused on scenario-based modeling as a possible
formalism for expressing override rules. There are other, related
modeling schemes, which could also be used in similar contexts. For
example, publish-subscribe is a related framework for parallel
composition, which shares many traits with
SBM~\cite{EuFeGuKe03}. Aspect oriented
programming~\cite{KiLaMeMaLoLoIr97} is another formalism that allows
to specify and execute cross-cutting program instructions on top of a
base application. Both of these approaches, however, do not directly
support specifying forbidden behavior, which appears quite useful for
specifying override rules. Additional behavior- and scenario-based models,
such as Brooks's subsumption architecture~\cite{Br86}, Branicky's
behavioral programming~\cite{Br99}, and LEGO Mindstorms leJOS
(see~\cite{Ar98}), all call for constructing systems from individual
behaviors. One advantage that SBM affords compared to these formalisms
 is that it is language-independent, has been implemented on top of multiple
 platforms, and can extend in a variety of ways the coordination and
 arbitration mechanisms used by these architectures.

The BIP formalism (behavior, interaction, priority) uses the notion of
glue for assembling components into a cohesive
system~\cite{BlSi08}. The goals that it pursues are similar to those
of SBM, although BIP's focuses mostly on correct-by-construction
systems --- while SBM is more geared towards  executing intuitively
specified scenarios, and resolving the constraints that they pose at
run-time.

\section{\uppercase{Discussion and Next Steps}}
\label{sec:conclusion}
\noindent

With the increasing use of DNNs in various systems, there is an urgent
need to ensure their safety, specifically by using override rules. We
argue here that progress can be made towards this goal by using
modeling schemes that model together the DNN and its override rules.
We propose to use scenario-based modeling for this purpose, show how
the basic scenario-based scheme can be extended to incorporate DNNs,
and demonstrate the approach on several examples.

Moving towards a more structured methodology for modeling override
rules raises the following question: as the number of override rules
and their sophistication increases, could they fully capture the
model's logic and render the DNN obsolete? We believe that the answer
is negative, as override rules often forbid some specific behavior,
but still rely on the DNN to prioritize among the remaining options. We
believe that an optimal approach is to combine a DNN component with
appropriately modeled override rules, while maintaining and enhancing
both components throughout the system's lifetime.

Our work to date is but a first step, which we plan to
extend. Specifically, we intend to work on
\begin{inparaenum}[(i)]
  \item customizing the idioms of SBM, or related techniques, to
    better suit integration with DNNs and guard them in more subtle
    ways; and
  \item leveraging the other advantages of SBM, specifically its
    amenability to verification and automated analysis, in proving
    the overall correctness of DNN-enhanced models.
  \end{inparaenum}
  In the longer run, we believe that work in this direction will lead to
  the creation of DNN-enabled systems that are more robust
  and easier to maintain and extend.

\section*{\uppercase{Acknowledgements}}

We thank Yafim (Fima) Kazak for his contributions to this project, and
the anonymous reviewers for their insightful comments.  The project
was partially supported by grants from the Binational Science
Foundation (2017662) and the Israel Science Foundation (683/18).

\bibliographystyle{apalike}

\end{document}